# Collaborative Personalized Web Recommender System using Entropy based Similarity Measure

Harita Mehta[1], Shveta Kundra Bhatia[2], Punam Bedi[3] and V. S. Dixit[4]

[1] Computer Science Department, Acharya Narender Dev College, University of Delhi,
New Delhi 110019, India

[2] Computer Science Department, Swami Sharaddhanand College, University of Delhi,
New Delhi 110036, India

[3] Computer Science Department, University of Delhi,
Delhi 110007, India

[4] Computer Science Department, Atma Ram Sanatam Dharam College, University of Delhi,
New Delhi 110010, India

**Abstract**

On the internet, web surfers, in the search of information, always strive for recommendations. The solutions for generating recommendations become more difficult because of exponential increase in information domain day by day. In this paper, we have calculated entropy based similarity between users to achieve solution for scalability problem. Using this concept, we have implemented an online user based collaborative web recommender system. In this model based collaborative system, the user session is divided into two levels. Entropy is calculated at both the levels. It is shown that from the set of valuable recommenders obtained at level I; only those recommenders having lower entropy at level II than entropy at level I, served as trustworthy recommenders. Finally, top N recommendations are generated from such trustworthy recommenders for an online user.

## 1. Introduction

A web user is usually surrounded by the large quantity of heterogeneous information available on the dynamic web platform. This information overload makes it crucial for the web user to access personalized information. Thus, there is a need for powerful automated web personalization tools for "Web Recommendation" [1] which is primarily aimed at deriving right information at right time. Web recommender systems analyses web logs in order to infer knowledge from the web surfer's sessions and thereby generate effective recommendations for the surfer. It has been observed that, web surfer prefers to visit a page that was visited by another likeminded person in the recent past. User based Collaborative web recommender systems have the same role as that of such human recommenders [8, 18]. In such systems, a user profile is a vector of items and their ratings, continuously appended as the user interacts with the system over a specified period of time. This user profile is compared with the profiles of other users in order to find overlapping interests among users. Thus, it generates recommendations based on inter user similarity. The idea to use the concept of inter user similarity is that if a user has agreed with his neighbors in the past, he will do so in the future also. Trustworthiness is amount of confidence on each other, which exist among such pair of inter similar users. We put forward that, trustworthiness can be derived using entropy. Recommendations generated from such trustworthy users are always preferred over recommendations generated from an unknown user [2].

The current generation of web recommender systems, still require further improvements in order to make recommendation method more effective [6]. One of them is "Scalability problem". In order to find users with similar tastes, these systems require data from a large number of users before being effective, and as well as require a large amount of data from each user. Thus, the computational resources required to find inter similar users become a critical issue. In this paper, we propose a method to select trustworthy recommenders from the list of similar users. It is assumed that similar users are valuable users. We put forward that, trustworthiness between similar users can be calculated on the basis of entropy existing between them. The step II of the proposed algorithm runs at two levels, thereby selecting only trustworthy recommenders in order to reduce computational resources which would be required at the time of generation of recommendations. Entropy [19] is the measure of inter user similarity that exists during recommendation generation process. It is expressed in terms of discrete set of probabilities as given in Eq. (1).

$$H(D(U_t, U_x)) = -\sum_{i=1}^{n} p(d_i) \log_2 p(d_i) \qquad (1)$$

where, $D(U_t, U_x)$ is the difference score rating between the target user $U_t$ and user $U_x$ for $n$ unique URLs and $p(d_i)$ is the probability density function of difference score rating. These probabilities depict the degree to which the target user $U_t$ is similar to user $U_x$. Lower the entropy, higher the degree of inter user similarity. The paper is organized in the following sections. Section 2, emphasizes on past research on similar work. Section 3, discusses the proposed model for collaborative web recommender system followed by experimental study in Section 4. Section 5, concludes the proposed work.

## 2. Related Work

In collaborative filtering approaches, the system requires access to the item and user identifiers [5, 11]. A simple approach in this family, commonly referred to as user based collaborative filtering [16], creates a social network of users who share same rating patterns with the target user. This network of users is based on the similarity of observed preferences between these users and the target user. Then, items that were preferred by users in the social network are recommended to the target user. Item based collaborative filtering [13], recommends such items to the target user that were preferred by those other users who preferred the same set of items that were preferred by the target user in the past. In many applications, collaborative recommender systems adapt their behavior to individual users by learning their tastes during the interaction in order to construct a user profile that can later be exploited to select relevant items. User's interest are gathered in an explicit way (such as asking user to rate an item on a scale , rank items as per favorite or choosing one item out of many) or implicit way (such as keeping track of item's user views, keeping the list of items purchased in past , analyze users social network & discover similar likes or dislikes.) It is preferred to work with rating data generated implicitly from user's actions rather than explicit collection [15]. Logs of web browsing or records of product purchases, are as an implicit indication for positive opinions over the items that were visited or purchased.

There have been many collaborative systems developed in the academia and in the industry. Some of the most important systems using this technique are group lens / Net perception [17], Ringo / Firefly [6], Tapestry [20], Recommender [8]. Other examples of collaborative recommended system include the book recommender system from amazon.com, the PHOAKS system that helps people find relevant information on WWW [14], and the Jester system that recommends jokes [12]. According to [11], algorithms for collaborative recommendation can be grouped into two general classes: memory based (heuristic based) and model based. Memory based collaborative filtering systems compare users against each other directly using correlation or other similarity measures such as scalar product similarity, cosine similarity and adjusted cosine similarly measure. Model based collaborative filtering systems derive a model from historical rating data and use it to make predictions. In the proposed work, we are concentrating on model based user collaborative filtering system. The researchers are trying to improve the prediction accuracy of generated recommendations. Recommendations generated by trustworthy users are preferred over recommendations generated by unknown web surfers [3, 4]. Trustworthiness is the level of satisfaction which the user gets from another user. This has originated an emergent need to measure inter user similarity with respect to trustworthiness among them. In collaborative web recommender systems, inter user similarity can be measured using information entropy which can reduce prediction error in these systems. Means Absolute Error (MAE) is applied to measure the accuracy of recommendations. Lower MAE values represent higher recommendation accuracy [7, 9, and 10]. In [10], by use of similarity measure using weighted difference entropy, it was shown that the quality of recommendation was improved together with reduced MAE. In [9], another similarity weighting method using information entropy was used and showed reduction in MAE and was found to be robust for sparse dataset. In [7], entropy based collaborative filtering algorithm provided better recommendation quality than user based algorithm and achieved recommendation accuracy comparable to the item based algorithm. In our research, the proposed model measures entropy at two levels of a user session to find trustworthy users and generate recommendations with their degree of importance [3, 4] only from such trustworthy users and serve as a means to reduce scalability problem that hampered traditional collaborative filtering techniques.

## 3. Collaborative Personalized Web Recommender System using Entropy based Similarity Measure

The architecture of a "Collaborative Personalized Web Recommender System using Entropy based Similarity Measure" is proposed in figure 1. In our study, online recommendations are generated for the demo version of the website available at http://www.vtsns.edu.rs. On the request of online user, top N recommendations are generated by the proposed web recommender system. The main components of this system are Interface Unit, Offline Unit and Online Recommendation Generator.

Online user and the recommender system are two basic entities in any recommendation generation process. Interface unit acts as an interface between these entities. It fetches click stream pattern (pages visited by the user) from the current session of online user. It sends the request to the online recommender generator where top N recommendations are furnished for the online user. Finally, the interface unit accepts the generated recommendations and passes it to the browser, so that these recommendations can be displayed for the online user during his/her current session.

Offline unit is the heart of the proposed recommender system. Creation of the knowledge base for online recommender generator rests on this unit. The processor of the offline unit takes web log of the demo site as input and generates recommendations in offline mode for the user patterns stored in the web log. The processor is the backbone of the offline unit and runs in three steps as discussed below.

### 3.1 Data Preparation (Step I)

Relevant user sessions in the form of Page View (PV) binary matrix are obtained from the raw web log file with the help of pre processing tools i.e. Sawmill [21]. Binary cell value as "1" in the matrix depicts that the page $P_m$ has been accessed in the session id $S_n$ whereas "0" depicts that the $P_m$ has not been accessed in the session id $S_n$. The PV matrix is split into Training PV matrix ($T_1$) and Test PV matrix ($T_2$); Training PV matrix ($T_1$) is further split into training level I matrix ($M_1$) and training level II matrix ($M_2$) which are required inputs for Step II of the processor.

### 3.2 Selection of Trustworthy Recommenders based on Entropy between the users (Step II)

This step can be broken down into two levels. At level I, Training level I matrix ($M_1$) is given as input. After initializing users, difference score is calculated between target user and all other users using Eq. (2).

$$D(U_t, U_x) = \left[ \left| PV_{(U_t, P_1)} - PV_{(U_x, P_1)} \right|, \cdots \left| PV_{(U_t, P_n)} - PV_{(U_x, P_n)} \right| \right] \quad (2)$$

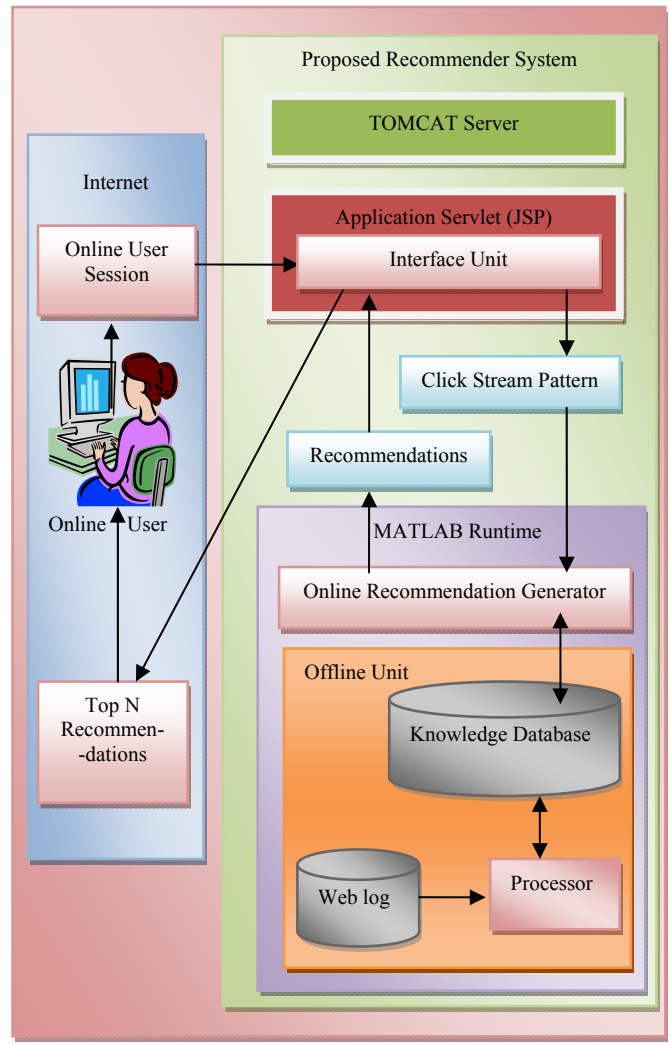

Fig. 1 Architecture of Proposed Recommender System

where, each term represents the page view status of user $U_k$ for page $P_n$. And, the absolute difference of the page view status for page $P_i$ of two users is considered as $(d_i)$ which is "0" when target user $U_t$ and user $U_x$ have both either viewed or have not viewed the page $P_i$. Parameter β introduced in Eq. (3) is a similarity threshold which is used to declare whether a web user is a valuable recommender for the target user or not.

$$DZeroCount(U_t, U_x) \geq \beta \times length(D(U_t, U_x)) \quad (3)$$

Here, we count the number of non-zero $(d_i)$ in each $(U_t, U_x)$ pair. If this number is greater than or equal to β times of total number of $(d_i)$ present, then we declare user $U_x$ to be a valuable recommender for target user $U_t$. Level I entropy $(E_I)$ among such pair of valuable recommenders is calculated using Eq. (1) and for each target user, list of valuable recommenders arranged in descending order of level I entropy is produced. This set of valuable recommenders for all the users and Training level II matrix ($M_2$) are given as input at level II. For such valuable users, level II entropy ($E_{II}$) is calculated using Eq. (1). Lower the entropy, higher the inter-user similarity. If level I entropy is less than level II entropy, then inter user similarity is more. It depicts that the interest of the user remains similar to that of the target user. So the user is considered as trustworthy user for the target user. For such pairs, actual entropy ($E_A$) is obtained using Eq. (4).

$$E_A(U_t, U_x) = (E_I(U_t, U_x) - E_{II}(U_t, U_x))/2 \qquad (4)$$

Finally, list of trustworthy recommenders arranged in descending order of actual entropy is produced because lower the entropy, higher the similarity. The algorithm is depicted in figure 2 and figure 3. Our approach reflects a solution to scalability problem in step II, by reducing computational resources; since it generates recommendations only from trustworthy recommenders.

3.3 Generation of Recommended Pages with their degree of importance (Step III)

Set of trustworthy recommenders for all users ($R_T$) obtained from Step II along with Page View matrix (PV matrix) prepared in Step I and Page visit frequency count (total number of users who have accessed that page) are given as input. Those pages which have not been visited by the target user, but have been visited by its trustworthy recommender, are considered as recommended page for the target user. Finally, evaluate degree of importance for generated recommendations using Eq. (5). Algorithm is depicted in figure 4.

$$DOI(U_t, P_{rec}) = (1 - E_c/T_c) \times F_c \qquad (5)$$

where, $T_c$ is the total number of trustworthy recommenders who have recommended the page $P_{rec}$ to the target user $U_t$, $F_c$ is the total number of users who have viewed the page $P_{rec}$ and $E_c$ is the total actual entropy value that the trustworthy users have assigned to the page $P_{rec}$. These recommendations generated by the processor in the offline mode act as knowledge base for the online recommendation generator. The knowledge base constructed by the offline unit consists of user click stream patterns and their recommended pages. Ongoing session information of the online user captured by the interface unit is given as input to the online recommendation generator. It matches the online partial click stream pattern of the online user with the partial click stream patterns of same length stored in knowledge base and finds trustworthy recommenders. Finally, Top N recommendations from the set of these trustworthy recommenders are provided to the interface unit. The algorithm is shown in figure 5.

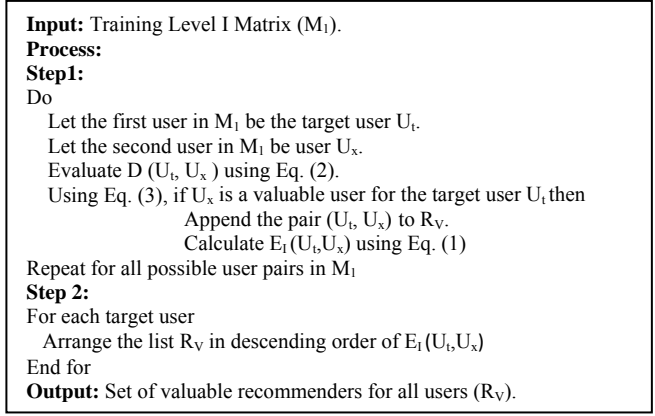

Fig. 2 Algorithm for Level I

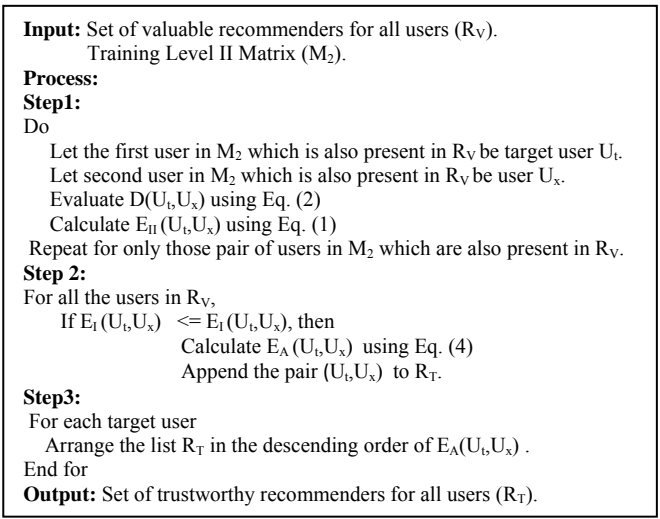

Fig. 3 Algorithm for Level II

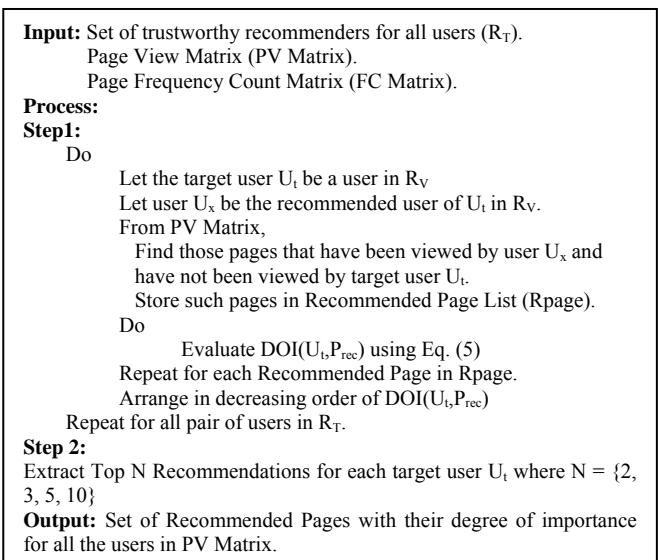

Fig. 4 Algorithm for Offline Unit

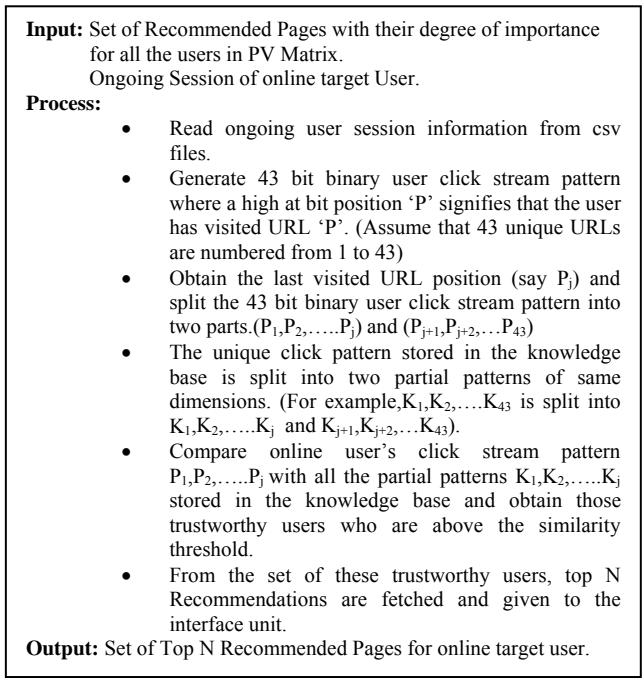

Fig. 5 Algorithm for Online Recommendation Generator

## 4. Experimental Study

### 4.1 Dataset

The demo version of the website (http://www.vtsns.edu.rs) was prepared using Microsoft front page. A prototype of the proposed system was implemented using MATLAB software [22] with TOMCAT server [24] on JAVA platform [23]. The internet platform was realized based on Java Server Pages (JSP) with Tomcat Server as servlet container. Here, Tomcat server acted as a container of the system servlet. The servlet itself was written in JSP and was run on Matlab software. The online target user with the help of web browser got connected with the server via internet provided by Tomcat server. The demo version of the website viewed by the online target user was displayed by the application servlet written in JSP. This servlet gathered the click stream of the online target user via web browser and sent it to the Matlab runtime library. The online recommendations generated were returned by Matlab to the application servlet which displayed them in the demo site via web browser. The experiment proceeded in a desktop PC environment consisting of Intel Core 2 Duo @ 3.00GHz and 2GB RAM. A web usage log file (http://www.vtsns.edu.rs/maja/vtsnsNov16) containing 5999 web requests to an institution's official website on November 16, 2009 was used as dataset. Sawmill processed these requests and grouped the hits into initial sessions based on the visitor id by assuming that each visitor contributes to a session. A session timeout interval of 30 minutes was considered for generating final sessions and sessions longer than 2 hours were eliminated. Page view count is the number of pages accessed by the user. Average page view count obtained from page view matrix was 5.4. So, we optimized our matrix by deleting those sessions that had visited less than 5 pages and deleted those URLs which were visited in only one or two sessions. Finally, we obtained 122 sessions with 43 unique URLs which was used as the input to verify the proposed recommendation generation process. Table 1 shows sample data of 5 users. We considered 80% of the dataset as training page view matrix and rest 20% as test page view matrix. Further, for calculating entropy at two levels, training page view matrix was split vertically with 22 pages at level I and rest 21 pages at level II. We assumed similarity threshold $\beta = 80\%$.

Table 1: Page View Matrix (sample data for 5 users)

| Page/User | P1 | P2 | …. | …. | P41 | P42 | P43 |
|---|---|---|---|---|---|---|---|
| U1 | 1 | 0 | …. | …. | 0 | 1 | 1 |
| U2 | 1 | 1 | …. | …. | 1 | 0 | 1 |
| U3 | 1 | 1 | …. | …. | 0 | 0 | 0 |
| U4 | 1 | 0 | …. | …. | 1 | 0 | 1 |
| U5 | 0 | 0 | …. | …. | 0 | 1 | 0 |

### 4.2 Results

Figure 6(a) obtained after running step I and II shows a graph depicting number of valuable and trustworthy recommenders. It can be clearly seen that the number of valuable recommenders obtained at level I are considerably decreased at level II to obtain number of trustworthy recommenders. For example, in case of user $U_{42}$, out of 36 valuable users only 28 users are trustworthy and in case of user $U_{43}$, the number reduced from 11 to 7.

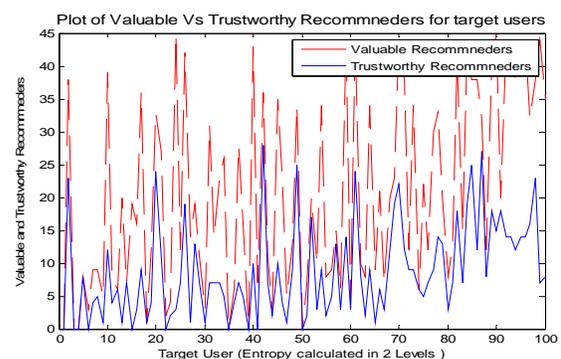

Fig. 6(a) Graph depicting Number of Valuable and Trustworthy Recommenders for 100 users

Figure 6(b) shows the priority of the trustworthy users for user $U_{43}$. Figure 7 shows plot of level I entropy and level II entropy for trustworthy users of target user $U_{43}$. It can be clearly visualized that, user $U_{97}$ has higher priority than user $U_{88}$ because the difference between level I entropy and level II entropy is lesser than that for user $U_{88}$. Finally, recommendations were generated at step III and Figure 8 depicts degree of importance of recommended pages for target user $U_{43}$.

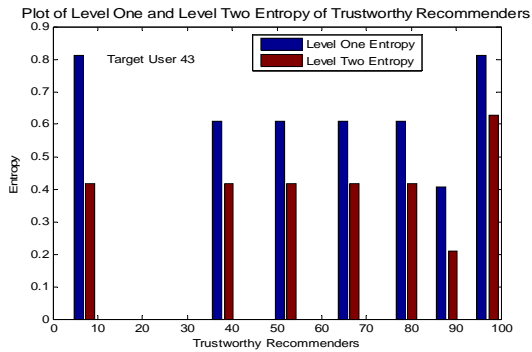

Fig. 6(b) Prioritized trustworthy users for user $U_{43}$.

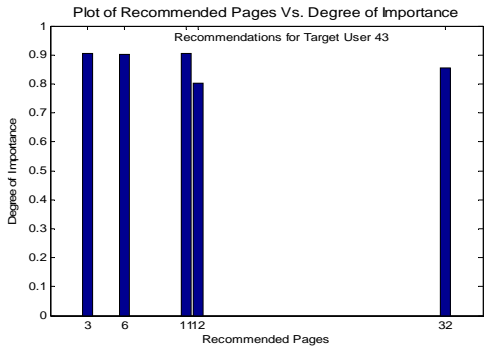

Fig. 7 Plot of Level I Entropy and Level II Entropy of Trustworthy Recommenders for the target user $U_{43}$.

Fig. 8 Plot of Recommended Pages vs. Degree of Importance for target User $U_{43}$.

We conducted a set of experiments to better understand how entropy calculated at levels improves the selection of trustworthy users. At step III, top N recommendations with their degree of importance was obtained where N= {2, 3, 5, 10}. To check the efficiency in offline mode, it was assumed that recommendations were generated after the target user has visited at least 6 URLs on the website. In order to find out similar users for the target user, similarity threshold β was set to 50% (i.e. at least 3 similar clicks). For this purpose, training page view matrix was split into visited training PV matrix (containing those 6 URLs already visited) and unvisited training PV matrix (rest of unvisited URLs). Similarly, test page view matrix was split into visited test PV matrix and unvisited test PV matrix of same dimensions. For each target user in visited PV test matrix, similar users were identified from visited PV training matrix. From the list of top N recommendations generated at step III, recommendations were obtained for these similar users and were stored in the predicted list. Finally, from the unvisited test PV matrix, actual pages viewed were found and stored in the actual list. In this experiment, we used Means Absolute Error (MAE), a statistical accuracy metric. Suppose, the set of entropy values predicted from the training set is {$p_1, p_2, \ldots p_n$}, and the corresponding set of actual entropy values from the test set is {$q_1, q_2, \ldots q_n$} then MAE is obtained using Eq. (6).

$$MAE = \frac{\sum_{i=1}^{n}\left(|p_i - q_i|\right)}{N} \quad (6)$$

where, $p_i$ is the predicted entropy, $q_i$ is the actual entropy, n is total number of URLs and N is total number of levels. Lower MAE values represent higher trust value between the pair of users because the interests of these users remain same throughout the session. MAE values obtained for Top N recommendation sizes are shown in table 2. The graph in figure 9 depicts that for all recommendation sizes, MAE values remained less than 0.5.

Table 2: MAE for Top N Recommendations

| Top N | Top 2 | Top 3 | Top 5 | Top 10 |
|---|---|---|---|---|
| Proposed System | 0.2114 | 0.2591 | 0.3245 | 0.4027 |

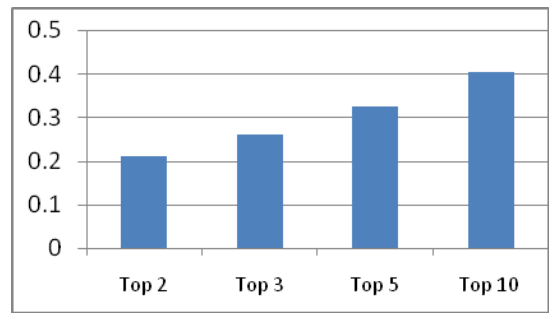

Fig 9 MAE for Top N Recommendations

To measure the quality of the proposed recommender system, two information retrieval measures, Precision and Recall were studied. Precision is the proportion of recommendations that are good recommendations and recall is the proportion of good recommendations that appear in top recommendations. Suppose, the set of URLs that are viewed by the target user are Relevant URLs and those URLs that are recommended by the recommender are Retrieved URLs, then precision ratio and recall ratio are obtained using Eq. (7a) and (7b) respectively.

$$\text{Precision} = \frac{|(\text{Relevant URLs}) \cap (\text{Retreived URLs})|}{|(\text{Retreived URLs})|} \quad (7a)$$

$$\text{Recall} = \frac{|(\text{Relevant URLs}) \cap (\text{Retreived URLs})|}{|(\text{Relevant URLs})|} \quad (7b)$$

One cannot achieve 100% Precision ratio or Recall ratio. So, we understand them relatively in relation to other systems. For this comparison, we prepared a single level entropy based recommender system (In introduction, we argued that algorithm will run in two levels at step II in order to generate recommendations only from trustworthy recommenders thereby reducing the required computational resources. To prove the statement, we implemented another Single Level Entropy based algorithm (SLE Web Recommender), in which the dataset was not divided into two sessions. It selected valuable users for a target user implicitly based on inter user difference score similarity obtained from page view matrix. Further, entropy for such valuable recommenders was calculated from the entire dataset. Similarity threshold was set to half of difference of maximum entropy and minimum entropy of the system. Those valuable recommenders who had entropy less than similarity threshold were considered as trustworthy recommenders. Finally, from such trustworthy recommenders, recommended pages with their degree of importance were obtained.) SLE web recommender was compared with our proposed web recommender system. Precision and Recall ratios recorded at various recommendation sizes is shown in table 3(a) and 3(b) respectively. Further, corresponding graphs are shown in figure 10(a) and 10(b). From this viewpoint, the measurements of our system showed better performance in both precision and recall ratios. It can be clearly seen that as the recommendation size increases, precision ratio decreases whereas recall ratio increases. For top 5 recommendation size, precision ratio marginally increased but recall increased almost 2.2 folds (i.e. from 24.1 % to 53.1%). Also, for top 10 recommendation size, recall increased almost 2.5 folds (i.e. from 24.1 % to 61.9%) and precision ratio marginally increased. Recall measures may be improved by increasing the recommendation size; however, it is best not to recommend too many items to users in order to avoid overloading. Choosing a proper recommendation size will be an appropriate topic for future studies.

Table 3(a): Precision Ratios

| Top N | Top 2 | Top 3 | Top 5 | Top 10 |
|---|---|---|---|---|
| SLE Web Recommender | 19.90% | 18.10% | 18.10% | 18.00% |
| Proposed Web Recommender | 30.30% | 27.10% | 24.50% | 22.10% |

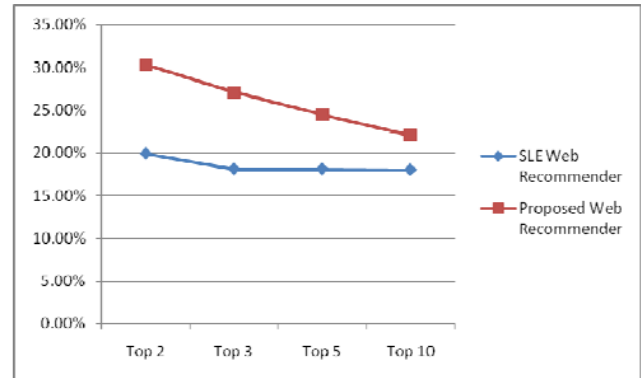

Fig 10(a) Precision Ratio for Top N Recommendations

Table 3(b): Recall Ratios

| Top N | Top 2 | Top 3 | Top 5 | Top 10 |
|---|---|---|---|---|
| SLE Web Recommender | 21.50% | 23.70% | 24.10% | 24.10% |
| Proposed Web Recommender | 30.10% | 38.70% | 53.10% | 61.90% |

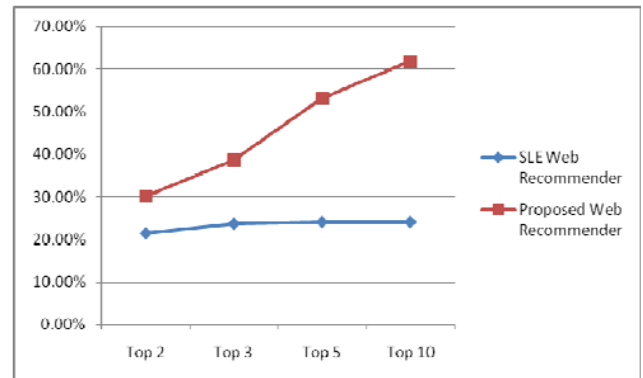

Fig 10(b) Recall Ratio for Top N Recommendations

After running steps I to III in offline mode, knowledge database was created. The database contained unique click patterns and their recommended pages. A demo version of the site available at http://www.vtsns.edu.rs was developed. Figure 11(a) shows the snapshot of the demo site. The snapshot of top N recommendations generated online is displayed in figure 11(b). Top N Recommendations were generated for an online user and similarity threshold β was set to 50%.

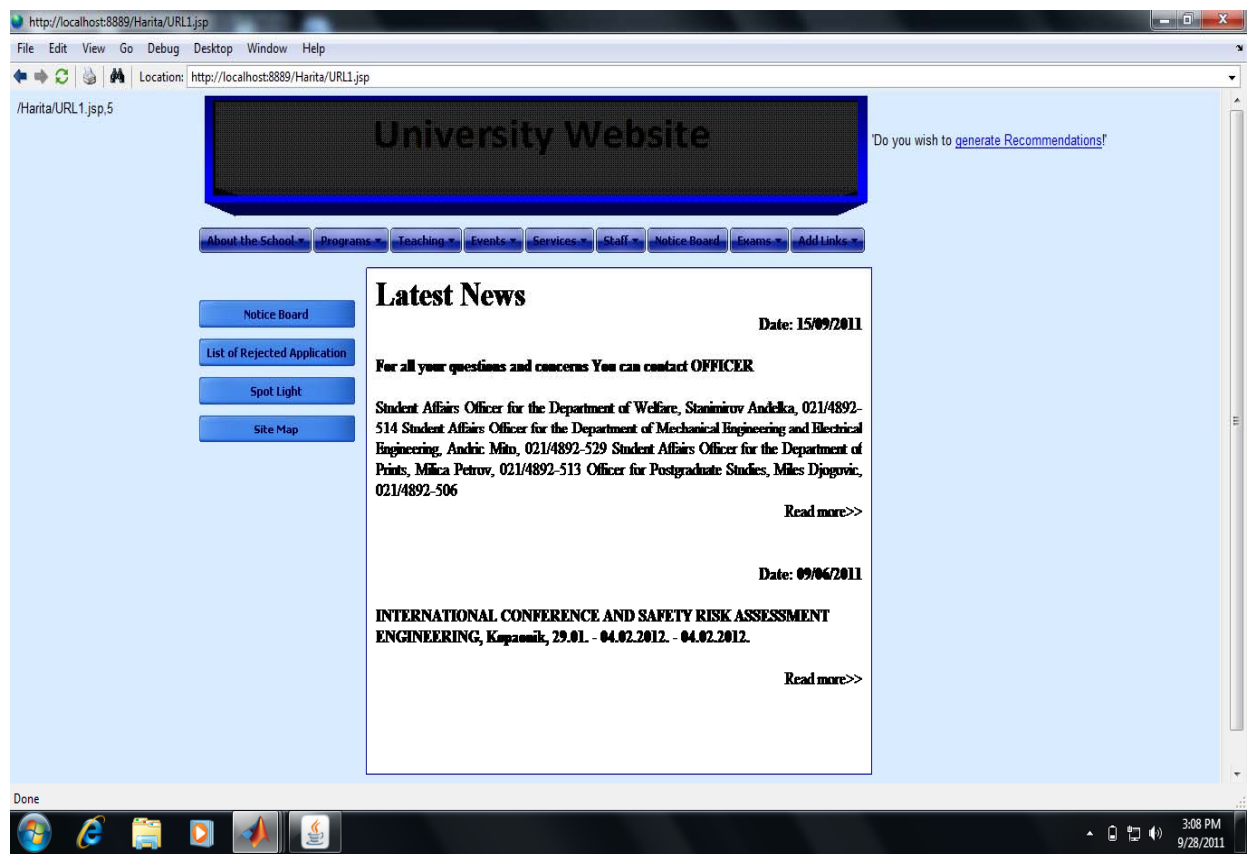

Fig. 11(a) Snap Shot of Demo Site

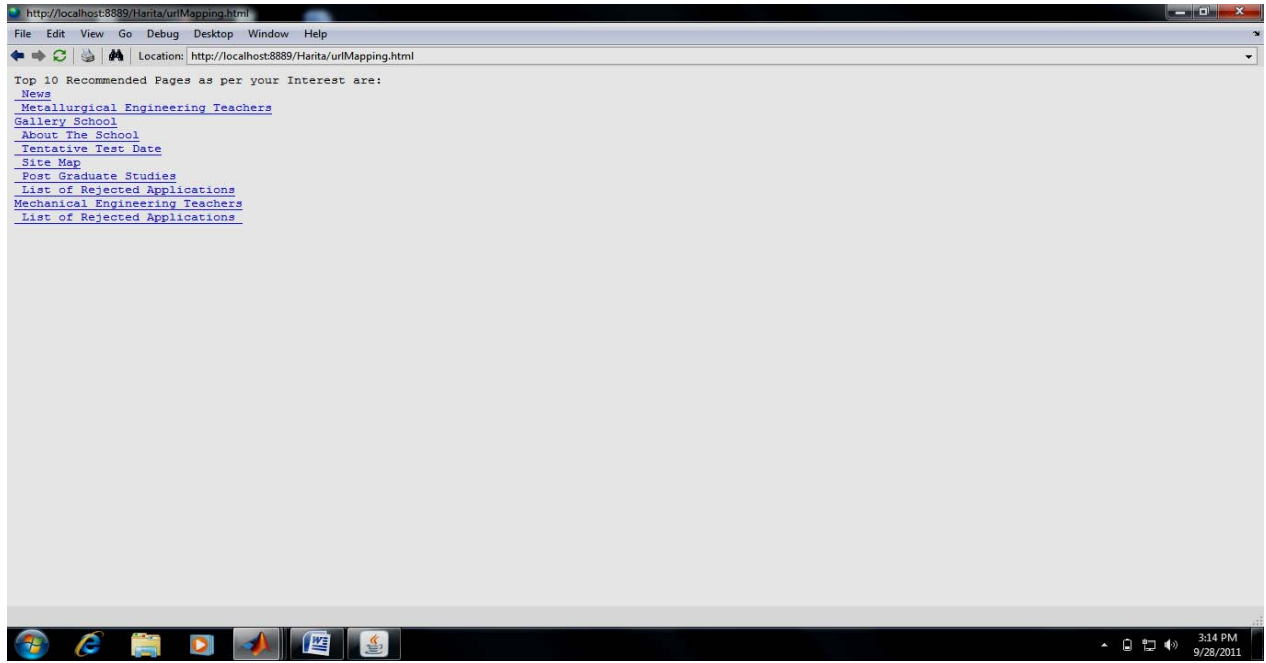

Fig. 11(b) Snap Shot of Top 10 Recommendations for an online user

## 5. Conclusions

The interest in the area of collaborative web recommender system still remains high because of the abundance of practical applications that demands personalized recommendations. In this paper, a "Collaborative Personalized Web Recommender System using Entropy based Similarity" is implemented in order to solve the problem of scalability. Traditionally, collaborative systems have relied heavily on inter user similarity based on difference score rating. We have argued that the difference score similarity on its own may not be sufficient to generate effective recommendations. Specifically, we have introduced the notion of entropy in reference to degree to which one might trust a specific user during recommendation generation. We have developed entropy based computational model which operated at two levels instead of single level. At both levels, recommenders were generated by monitoring entropy between similar users based on difference score rating. We have described a way to suppress the generation of recommenders who were valuable but not trustworthy. We found that the use of entropy at two levels had a positive impact in solving scalability problem. Top N recommendations were generated and MAE was found to be less than 0.5 for all recommendation sizes. As the recommendation size increased, Precision ratio decreased and recall ratio increased. Precision and recall for top N recommendations were found to be better when compared with single level entropy based web recommender system.

**Ms. Harita Mehta** is a Research Scholar and working as an Assistant Professor in the Department of Computer Science, Acharya Narender Dev College, University of Delhi. Her research area is Web Recommender Systems and is currently pursuing PhD under Dr. V.S. Dixit from Department of Computer Science, University of Delhi.

**Ms. Shveta Kundra Bhatia** is a Research Scholar and working as an Assistant Professor in the Department Of Computer Science, Swami Sharaddhanand College, University of Delhi. Her research area is Web Usage Mining and is currently pursuing PhD under Dr. V.S. Dixit from Department of Computer Science, University of Delhi.

**Dr. Punam Bedi** received her Ph.D. in Computer Science from the Department of Computer Science, University of Delhi, India in 1999 and her M.Tech. in Computer Science from IIT Delhi, India in 1986. She is an Associate Professor in the Department of Computer Science, University of Delhi. She has about 25 years of teaching and research experience and has published more than 110 research papers in National/International Journals/Conferences. Dr. Bedi is a member of AAAI, ACM, senior member of IEEE, and life member of Computer Society of India. Her research interests include Web Intelligence, Soft Computing, Semantic Web, Multi-agent Systems, Intelligent Information Systems, Intelligent Software Engineering, Software Security, Intelligent User Interfaces, Requirement Engineering, Human Computer Interaction (HCI), Trust, Information Retrieval, Personalization, Steganography and Steganalysis.

**Dr. V. S. Dixit** is working as senior Assistant Professor in the Department Of Computer Science, AtmaRam Sanatam Dharam College, University of Delhi. His research area is Queuing theory, Peer to Peer systems, Web Usage Mining and Web Recommender systems. He is currently engaged in supervising the research scholars (Ms Harita Mehta and Ms Shveta Kundra Bhatia) for PhD. He is Life member of IETE.